%Paper: hep-th/9506058
%From: Tomoyasu Ichihara <tomo@th.phys.titech.ac.jp>
%Date: Fri, 9 Jun 1995 16:11:42 +0900

\input harvmac
\input epsf
\overfullrule=0pt
\def\caption#1#2{\smallskip\noindent
{\rm#1}{\it#2}\medskip\noindent}

\def\nl{\hfil\break}
\def\tit{\medskip\centerline{\it Department of Physics}
\centerline{\it Tokyo Institute of Technology}
\centerline{\it Oh-okayama, Meguro-ku}
\centerline{\it Tokyo { 152}, JAPAN}
\vskip .3in}
\def\ko{\medskip\centerline{\it
  Department of Physics, Hiyoshi Campus}
 \centerline{\it Keio University}
  \centerline{\it Hiyoshi, Kouhoku-ku}
  \centerline{\it Yokohama { 223},   JAPAN}
\vskip.3in}

\def\npb{{ \sl Nucl. Phys. }}
\def\prc{{ \sl Phys. Rep. }}
\def\prd{{ \sl Phys. Rev. }}
\def\prl{{ \sl Phys. Rev. Lett. }}
\def\plb{{ \sl Phys. Lett. }}

\font\cmss=cmss10 \font\cmsss=cmss10 at 7pt
\def\IZ{\relax\ifmmode\mathchoice
{\hbox{\cmss Z\kern-.4em Z}}{\hbox{\cmss Z\kern-.4em Z}}
{\lower.9pt\hbox{\cmsss Z\kern-.4em Z}}
{\lower1.2pt\hbox{\cmsss Z\kern-.4em Z}}\else{\cmss Z\kern-.4em Z}\fi}
\def\undertext#1{\vtop{\hbox{#1}\kern 1pt \hrule}}
\def\expec#1{\left\langle#1\right\rangle}
\def\half{{1\over2}}
\def\c#1{{\cal{#1}}}
\def\slash#1{\hbox{{$#1$}\kern-0.5em\raise-0.1ex\hbox{/}}}
\def\dirac{\hbox{$\partial$\kern-0.5em\raise0.3ex\hbox{/}}}
\def\dslash{\hbox{$\partial$\kern-0.5em\raise0.3ex\hbox{/}}}
\def\Dirac{\hbox{{\it D}\kern-0.52em\raise0.3ex\hbox{/}}}
\def\Dslash{\Dirac}
\def\pint{P\!\!\!\!\int}
\def\kslash{\hbox{{\mit k}\kern-0.4em\raise0.3ex\hbox{/}}}
\def\pslash{\hbox{{\it p}\kern-0.48em\raise-0.3ex\hbox{/}}}
\def\qslash{\hbox{{\mit q}\kern-0.5em\raise-0.3ex\hbox{/}}}
\def\gsim{\mathrel{\raise.3ex\hbox{$>$\kern-.75em\lower1ex\hbox{$\sim$}}}}
\def\lesssim{\mathrel{\raise.3ex\hbox{$<$\kern-.75em\lower1ex\hbox{$\sim$}}}}

\def\nc{N_c}

\def\be#1{\beta_{#1}}
\def\bb#1{\gamma_b}
\def\bbf#1{\gamma_f}
\def\sgn{\,{\rm sign}\,}

\def\lm{\lambda_-}

\def\tvp{\tilde\varphi}

\def\ppint{\pint_0^1\!\!\!{dy\over(y-x)^2\,}}
\def\hqn#1{\varphi^{hq}_{#1}}
\def\sumk#1{\sum_{#1=1}^K}
\def\intk{\int\!{d^2k\over(2\pi)^2}}
\def\xzz{x_0}
\nopagenumbers\abstractfont\hsize=\hstitle\rightline{TIT/HEP--263}
\vskip 1in\centerline{\titlefont
1+1 dimensional QCD
with fundamental bosons and fermions}
\abstractfont\vskip 1truecm\pageno=0
\centerline{Kenichiro Aoki\footnote{$^1$}{\hskip-1mm%
email:{\tt~a00500@cc.hc.keio.ac.jp}}}
\ko
\centerline{ Tomoyasu Ichihara\footnote{$^2$}{\hskip-1mm%
email:{\tt~tomo@th.phys.titech.ac.jp}}}
\tit
\medskip
\centerline{\bf Abstract}
We analyze the properties of mesons in 1+1 dimensional QCD with
bosonic and fermionic ``quarks'' in the large $\nc$ limit.
We study the spectrum in detail and show that it is impossible
to obtain massless mesons including boson constituents in this
model.
We quantitatively show how the QCD mass inequality is realized
in two dimensional QCD.
We  find that the mass inequality is close to
being an equality even when the quarks are light.
Methods for obtaining the properties of ``mesons'' formed from boson
and/or fermion constituents are formulated in an explicit
manner convenient for further study.
We also analyze how the physical properties of the
mesons such as confinement and asymptotic freedom are realized.
\Date{}
\newsec{Introduction}
In confining gauge theories, physically observable
particles at low energies have no gauge charge and are bound
states of the charged matter that appear in the gauge theory
Lagrangian.
A prototype of such a theory is QCD wherein the
gauge group is SU$(3)$ and the charged matter fields are  vector
fermions in the fundamental representation.
Several natural generalizations of QCD exist;  we may use
gauge groups other than SU$(3)$, we may use chiral fermions or
we may choose representations more complicated than the
fundamental representation for the matter fields.
In this case, we cannot put in too many or too large
representations if we want to preserve asymptotic freedom.
Another generalization is to consider boson constituents as well
as fermion constituents.

It is this last, as well as the first  generalization in two
dimensions in the large
$\nc$ limit that we shall investigate in this work.
The possibility of boson constituents arise necessarily in
several contexts, such as technicolor with multiple stages of
symmetry breaking\ref\TC{For a review on technicolor, see for
instance,  E.~Farhi, L.~Susskind, \prc{\bf 74} (1981) 277
}, QCD or technicolor with supersymmetry\ref\STC{For reviews,
see, for instance, D. Amati et al, \prc {\bf 162} (1988) 169\nl
N.~Seiberg, TASI lectures (1994) },
 so called bosonic technicolor\ref\BT{E.H. Simmons, \npb{\bf
B312} (1989) 253\nl
S. Samuel, \npb{\bf B347} (1990) 625} as well as the Standard
Model\ref\AF{L.~Abbott, E.~Farhi, \plb{\bf 101B} (1981) 69,
\npb{\bf B189} (1981) 547}.
When both boson and fermion constituents exist in the theory,
``meson'' states of both Bose and Fermi statistics may arise.
In general, it is difficult to derive the properties of the
bound states from first principles.
By using the large $\nc$ limit in two dimensions, we may
analyze the properties of these meson states concretely.

Two dimensional QCD in the large $\nc$ limit has greatly
contributed to our current understanding of the gauge theory
dynamics by providing us with a model where the properties are
explicitly calculable analytically.
Also, two dimensional QCD has been used to test the validity of
various approaches and approximation schemes applied to QCD.
The model was first solved by {}'t Hooft \ref\THOOFT{
 G.~{}'t Hooft, \npb{\bf B72} (1974) 461, {\bf B75} (1974) 461}
and some further physical properties such as some current matrix
elements,  the asymptotic freedom of mesons were
studied in some subsequent works \ref\CCG{C.G.~Callan, N.~Coote,
D.J.~Gross, \prd{\bf D13} (1976) 1649}%
\ref\EINHORN{M.B. Einhorn, \prd{\bf 14D} (1976) 3451, \prd {\bf
15D} (1977) 3037}.
The formulation was extended to include  boson--boson
bound states \ref\BP{W.A.~Bardeen,
P.B.~Pearson, \prd{\bf D14} (1976) 547;\nl
M.B.~Halpern, P.~Senjanovic, \prd{\bf D15} (1977) 1655}%
\ref\SHEI{S-S. Shei, H-S. Tsao, \npb{\bf B141}
(1978) 445}
and boson--fermion bound
sates\ref\KABF{K. Aoki, \prd{\bf D49} (1994) 573}.
Mesons made only from fermionic quarks or the bosonic quarks
obey Bose statistics but the boson--fermion bound state obeys
Fermi statistics.
(Hereafter, often referred to as ff, bb and bf cases.)

In this work, we shall extend the investigation of the physical
properties of the mesons made from fermions and generalize the
results  to the mesons made from bosons only and bosons and
fermions.
The results will enable us to compare the three cases and see
the differences and similarities that arise between mesons made
from constituents of various statistics.
The spectrum of mesons is investigated both analytically and
numerically. We will study the case when the
quarks are heavy analytically. For mesons involving light
quarks, we establish a number of results analytically and
further analyze the problem numerically.
The numerical approach  will be formulated explicitly in all the
three ff, bf and bb cases which should be useful for
further study.

We use these results to see how the QCD mass
inequality\ref\INEQ{D. Weingarten, \prl{\bf51} (1983) 1830\nl
E.~Witten, \prl{\bf 51} (1983) 2351}
\ref\NUSSINOV{S. Nussinov, \prl{\bf 51} (1983) 2081}
applies to two dimensional QCD in the large $\nc$ limit.
Given two different types of quarks, the QCD mass inequality
states that the meson made from the same quarks is on average
lighter than  the meson made from different quarks.
This non--trivial inequality, however, does not
tell us by how much these meson masses differ, a question which
we are able to answer analytically in some cases and numerically
for all quark masses.
Also, while there is no reason to doubt this important
inequality, as was pointed out in the original articles
themselves, the inequality is not completely rigorous.
We find it satisfying to be able to study how the inequality is
realized in a simplified version of QCD.
To our knowledge, the QCD mass inequality has not been previously
shown to be satisfied  in such a concrete manner.
Here, the inequality is applied to ff, bf and bb cases. Except
in the bf case, the inequality is necessarily an equality when
the constituent quark masses are the same. In the bf case, such
needs not be the case and indeed we find that it is always a
strict inequality in the bf case.

The paper is organized as follows. In section 2, we briefly
review the integral equation satisfied by the wave function
of the meson  in the large $\nc$ limit, partly to fix the
notation.
In section 3, we analytically study the static properties of
mesons when the quark masses are large.
We further formulate the bound state equation as a linear
eigenvalue problem for general values of the quark masses and
analyze the problem numerically.
In sect 4, we analyze some matrix elements and see how
confinement and asymptotic freedom is realized in the bf case.
We conclude with a brief discussion.
\newsec{Wave function of mesons}
In this section, we briefly summarize the equations meson wave
functions satisfy and some basic properties of the solutions
in the ff\THOOFT, bf\KABF\ and bb\SHEI\  cases.
The classical Lagrangian of QCD coupled to fermions and  bosons
is
\eqn\lag{-\c L={1\over4}\tr(F_{\mu\nu}^2)
+\sum_f\overline\psi_f\left(\Dslash+m_f\right)\psi_f
+\sum_b\left(\left|D\phi_b\right|^2 +m_b^2\left|\phi_b\right|^2\right)
}
Both fermions and bosons are in the fundamental representation.
We shall refer to the fields $\psi_f,\phi_b$ as ``quarks''.
We fix the gauge to the light--cone gauge $A_-=A^+=0$,
where $a^\pm=a_\mp\equiv (a^1\pm a^0)/\sqrt2$.
Light--cone gauge has the advantage
that there are no gluon self interactions in 1+1
dimensions.

We take the large $\nc$ limit by letting $\nc$ go to infinity
keeping $g^2\nc$ and the quark masses to be of $\c O(1)$.
There are contributions of $\c O(1)$ to the quark propagators
which compete with the classical contribution.
These contributions  may be incorporated by solving the
Schwinger--Dyson equations recursively to obtain
the full propagators as follows:
\eqn\props{\eqalign{
S(p;m) &= \left[i\pslash+i{g^2\nc\over2\pi}
   \left({\sgn(p_-)\over\lm}-{1\over p_-}
  \right)\gamma^++m\right]^{-1}\cr
&=
\left[-i\pslash-i{g^2\nc\over2\pi}\left({\sgn(p_-)\over\lm}
-{1\over p_-}\right)\gamma^++m\right]\,D(p;m)
\cr
D(p;m) &=\left[p^2+m^2+{g^2\nc\over\pi}
        \left({|p_-|\over\lm}-1\right)\right]^{-1}\cr
}}
We note that the quantum corrections to the mass is identical
both for the fermionic and the bosonic quarks.
In this work, we will use an infrared cutoff $\lm$ which seems
to be more convenient for deriving the physical properties of
the mesons.
Other infrared regularizations may be used, but of course do
not affect physical results\CCG\EINHORN.

A meson is formed as a quark anti--quark bound state and its
wave function satisfies
\eqn\bseq{
\mu^2\tvp(x) =\left({\be a-1\over x}+{\be b-1\over1-x}\right)\tvp(x)
  -\ppint \tilde U(x,y)\tvp(y)\equiv\tilde H\tvp(x)}
where
\eqn\kdefeq{\tilde U(x,y)\equiv\cases{1&ff\cr(x+y)\over2x&bf\cr
        {(x+y)(2-x-y)\over 4x(1-x)}&bb\cr}}
Here, $\be {f,b}\equiv m_{f,b}^2\pi/ (g^2\nc)$, namely the mass
squared of the quarks counted in the units of the QCD scale.
These equations may be obtained by summing the graphs of the
ladder type.
Here and below, we refer to the solution of the Bethe--Salpeter
equation, $\tvp(x)$, as the wave function of the meson, in
analogy with the non--relativistic case and in
accordance with the previous literature.
$x\equiv p_-/r_-$ is the momentum fraction carried by the quark
and $1-x$ by the anti--quark in the infinite momentum frame.
In the $1/\nc$ expansion, sea quarks are suppressed by $1/\nc$
and gluons have no physical degrees of freedom so that the
all the momentum of the meson is carried by the quark and
the anti--quark.
$\pint$ denotes the principal part integral defined by
\eqn\ppintdef{\pint\!\! dx\,
f(x)\equiv\half\int\!\!dx \biggl[f(x+i\epsilon)
+f(x-i\epsilon)\biggr]_{\epsilon\rightarrow0}}
In the bf and the bb cases, the above meson equations
are not Hermitean with respect to the standard
measure $\int_0^1\!\!dx$ and we shall often use the conjugated
equation
\eqn\hermeq{
\mu^2\varphi(x)=\left({\be a-1\over x}+{\be b-1\over1-x}\right)\varphi(x)
  -\ppint
   %{\!\!dy\over(y-x)^2}\,
   U(x,y)\varphi(y)\equiv H\varphi(x)
}
\eqn\hermkdef{
  U(x,y)=\cases{1&ff\cr(x+y)\over2\sqrt{xy}&bf\cr
        {(x+y)(2-x-y)\over 4\sqrt{x(1-x)}\sqrt{y(1-y)}}&bb\cr}
}
The appropriate conjugation is
\eqn\conjug{\varphi(x)=\tvp(x)\times\cases{1&ff\cr
  \sqrt x&bf\cr\sqrt{x(1-x)}&bb\cr}}

A formula we find useful is
\eqn\useful{\eqalign{
 \left(\phi,H\psi\right)&=\int_0^1\!\!dx\,
 \left({\be a\over x}+{\be b\over 1-x}
 +J(x)\right)
 \overline{\phi(x)}\psi(x)\cr
  &\qquad+\half\int_0^1\!\!dx\,\int_0^1\!\!{dy\over(y-x)^2}\,
  U(x,y)(\overline{\phi(x)}-\overline{\phi(y)})
 (\psi(x)-\psi(y))\cr}}
where
\eqn\jdefeq{J(x)\equiv\cases{0&ff\cr
   -{1\over 1-x}\left({1\over x}-{1\over\sqrt x}\right)&bf\cr
   -{1\over x(1-x)}+{\pi\over4\sqrt{x(1-x)}}&bb\cr}}
Note that the expression no longer involves the principal part
integral.
{}From this equation, it  immediately follows that  in the ff case
when the  bare quark masses are positive ($\be a,\be b\geq0$),
the meson mass is positive\THOOFT.
In the bf, bb cases, it is clear that the meson mass is
positive when $\be a,\be b>1$.
Another property that may be obtained from this formula is that
for the same quark--antiquark masses, the mass of the meson is
the lightest for the bb case, heaviest in the ff case.
Here and below, we often refer to the bare masses of the quarks
$m_{f,b}$ as quark masses.
It should perhaps be commented here that the quark masses
$m_{f,b}$, or equivalently $\be {f,b}$, are not directly
physically observable and that they receive quantum
corrections.
However, we note that the quantum corrections to both the
fermionic and
the bosonic quark are identical so that it is consistent
to compare the quark masses on the same footing.

The wave function vanishes at the boundary as
$\varphi(x)\sim x^{\gamma_{f,b}}$ where $x$ denotes the momentum
fraction of the fermionic or the bosonic constituent.
$\gamma_{f,b}$ is determined by the equations
\eqn\bbeq{
  \cases{
    \be f-1+\pi\bbf2\cot(\pi\bbf2)=0& f \cr
     \be b-1-\pi\bb1\tan(\pi\bb1)=0& b\cr
  }
}

Since the meson equations are Hermitean, the spectrum is real.
The widths of the mesons are of $\c O(N_{\rm flavors}/\nc)$ and
are suppressed in the large $\nc$ limit.
For higher mass states, the approximate wave functions and
the respective masses are
\eqn\regge{
   \varphi(x)\simeq\sqrt2\sin\pi kx,\quad
   (\hbox{\rm meson mass})^2\simeq g^2\nc \pi k,\qquad
        k=1,2,3\ldots}
resulting in a linear Regge type trajectory for large $k$.
Here, $k$ labels the $k$--th lightest meson bound state formed from the
quark--antiquark pair.
We note that  the relativistic effects are important in this
problem, even qualitatively; in the non--relativistic case, the bound
state mass squared behaves as $k^{2/3}$ rather than linearly.
\newsec{Static properties of mesons}
In this section, we
obtain and analyze the static properties of mesons using both
analytical and numerical methods.
\subsec{Heavy quarks}
When the quark masses are heavy, we may use the variational
method to obtain the meson wave function analytically.
This method was previously applied to the ff case when the quarks
and anti-quarks have  equal mass \CCG.
Here, we extend the analysis to include the case when the
quark and the anti-quark have different
masses and further generalize this to the bf and bb cases.
We  refer to the ``heavy quark'' when the quark
mass is large compared to the QCD scale,
($m\gg g\ \Leftrightarrow\ \beta\gg1$).

Using the trial function
\eqn\trialfn{\hqn0(x)=\left(c\over\pi\right)^{1/4}
        e^{-c(x-\xzz)^2/2}}
we obtain
\def\www{J(\xzz)}
\eqn\trialres{(\varphi,H\varphi)=
{\beta_a\over \xzz}+{\beta_b\over 1-\xzz}+
  \left({\be a\over \xzz^3}+{\be b\over (1-\xzz)^3}\right){1\over2c}
       + \sqrt{\pi c}+\www+\c O(\beta c^{-2})}
Varying with respect to $\xzz$ and $c$ we obtain
\eqn\aceq{\eqalign{
  \xzz&={\sqrt {\be a}\over\sqrt{\be a}+\sqrt{\be
  b}}\left(1+\c O(c^{-1})\right)={m_a\over m_a+m_b}
  \left(1+\c O(c^{-1})\right)\cr
  c&=(\pi\be a\be b)^{-1/3}\left(\sqrt{\be a}
          +\sqrt{\be b}\right)^{8/3}
        \left(1+\c O(c^{-1})\right)\cr}}
Consequently,
\def\wws{w_s}
\eqn\mesonmass{
  \mu=(\sqrt{\be a}+\sqrt{\be b})\left[
    1+{3\over4}\pi^{1/3}(\be a\be b)^{-1/6} (\sqrt{\be a}
    +\sqrt{\be b})^{-2/3}+\wws\right]
 }
where
\eqn\wwsdefeq{
  \wws=\cases{0&ff\cr
    -{1\over2\sqrt{\be b}}\left({1\over\sqrt{\be a}}-
   {1\over\be a^{1/4}(\sqrt{\be a}+\sqrt{\be b})^{1/2}}\right)&bf\cr
   -{1\over2\sqrt{\be a\be b}}+{\pi \over4(\be a\be b)^{1/4}
  (\sqrt{\be a}+\sqrt{\be b})}&bb\cr}}
Here, we have obtained a clear physics picture of the meson as
 a bound state of the quark and anti--quark with the momentum
distributed in proportion to their masses.
As the $q,\overline q$ masses become larger, the momentum
distribution becomes narrower as $\beta^{-1/3}$   and this
simple constituent quark picture becomes more accurate.

Since we are considering different masses for the quark and the
anti--quark, it may seem we should analyze the effect of
including a variational function that is asymmetric with respect
to the center of the wave function, such as
$(x-\xzz)\exp-c(x-\xzz)^2/2$.
It can be immediately shown that to the order we considered
above, including this function in the variational problem does
not change the results at all.

To systematically obtain the wave function when the
quark masses are heavy
we may use an orthonormal  basis for the
wave functions
\eqn\basishq{\hqn n\equiv\left(c\over\pi\right)^{1/4}
        {1\over\sqrt{n!}} H_n(\sqrt{2c}(x-\xzz)) e^{-c(x-\xzz)^2/2}}
where $H_n$ is the $n$--th Hermite polynomial defined by
\eqn\hermitedef{e^{tz-t^2/2}=\sum_{n=0}^\infty
         {t^n\over n!}H_n(z)}
To determine $\xzz,c$ to leading order, as in \aceq, we need the
matrix elements of the Bethe--Salpeter equation to $\c
O(\beta^{1/3})$.
To this order, the statistics is unimportant.
In all the three ff, bf and bb cases, the matrix elements of $H$
are
\eqn\hqham{\eqalign{
  &\left(\hqn m, H\hqn n\right)
  =\cr
  &\ \cases{{\be a\over \xzz}+{\be b\over 1-\xzz}+
  {(2n+1)\over 2c}\left({\be a\over \xzz^3}+
   {\be b\over (1-\xzz)^3}\right)&\cr
  \qquad+\sqrt{\pi c}\left[1-\sum_{k=1}^n(-1)^k{n!(2k-3)!!\over
      (k!)^2(n-k)!}\right]
  +\c O(1)
  &$m=n$\cr
  {1\over {(2c)^{(m-n)/2}}}  \sqrt{m!\over n!}
  \left[
  {\be a\over \xzz^{m-n+1}}+{\be b\over(1-\xzz)^{m-n+1}}
  \right]&\cr
  \qquad-\sqrt{\pi c}\sum_{k=0}^n(-1)^{n-k}
  {\sqrt{m!n!}(m+n-2k-3)!!\over k!(n-k)!(m-k)!}
  +\c O(1)
  &$m>n,m-n\equiv0$ mod 2\cr
  {1\over {(2c)^{(m-n)/2}}}  \sqrt{m!\over n!}
  \left[-{\be a\over \xzz^{m-n+1}}+{\be b\over(1-\xzz)^{m-n+1}}
   \right ]
  +\c O(1)
  &$m>n,m-n\equiv1$ mod 2\cr}\cr}}
The one dimensional variational problem just using $\hqn0(x)$
is not quite enough to determine the lowest meson mass to the
order in $\beta^{-2/3}$ given above in \mesonmass.
It turns out, however,  that only when we consider trial
functions $\hqn n(x)$, with $n\geq4$ that the meson mass is
affected.
When we include the variational functions with
$n\geq4$, we find that the physical effect is to
decrease the numerical coefficient of the second term in the
square brackets of \mesonmass\ by less than
1\%. The last term, $\wws$,  is not affected at all.
The smallness of the corrections is not surprising; while the
contribution of more
complicated variational functions are not suppressed by powers
of $\beta^{-2/3}$, they are suppressed numerically just as the
contribution from the higher excited states to the ground state
energy are suppressed in perturbation theory in quantum
mechanics. This suppression is
strong since there is a contribution to the meson mass only when
we include variational functions with $n\geq4$.
A couple of comments are in order; first, the boundary
conditions at $x=0,1$ are not exactly satisfied in this
approach, so that the expansion is not completely systematic.
However, since the value of the wave function at the boundaries
is exponentially suppressed  as  $\exp -{\rm
const.}\times\beta^{2/3}$,
the approximation should be good when the quarks are heavy.
Second, the growth of the matrix elements with the increase in
the dimension of the variational space as seen in \hqham\
indicates that the expansion
in $\beta^{-2/3}$ is an asymptotic series.
This situation is quite common in quantum theories
\ref\LO{J.C.~Le~Guillou, J.~Zinn-Justin, (eds.),
{\sl``Large order perturbation theory"}, North Holland (1990)}.
Further analyzing the problem through this variational approach
does not change the physics picture obtained above
and we shall not pursue this further here.
\subsec{Variational methods using powers of the momentum fraction}
An effective method, both analytically and numerically,  for
obtaining approximate  solutions to the
integral equation for the meson wave function, \bseq,\hermeq\ is
to reduce  the problem to a finite dimensional diagonalization
problem.
An example of a variational approach particularly effective in
the case of heavy quarks  was given above.
When the quark masses are light, a variational scheme that is
effective is to use a basis
\def\basis#1{v_{#1}}
\eqn\polbasis{
  \basis{2k}=\left[x(1-x)\right]^{\gamma+k},\qquad
  \basis{2k+1}=\left[x(1-x)\right]^{\gamma+k}(1-2x)
  ,\qquad k = 0,1,2,\ldots
        }
This scheme was previously used in the ff case\ref\BPL{W.A. ~Bardeen,
R.B. Pearson, E. Rabinovici, \prd{\bf 21} (1980) 1037} and
more recently  in the investigations of the Schwinger model
\ref\SMODEL{J. Schwinger, \prd{\bf 128} (1962) 2425\nl
Y. Mo, R.J. Perry, {\sl Journal of Comp. Phys.} {\bf
108} (1993) 159\nl
K. Harada et al, \prd{\bf D49} (1994) 4226 and references therein.}.
Below, we treat the ff, bb cases when the quark and the
anti--quark masses are identical ($\be a=\be b=\beta$).
In this case, the integral equation \hermeq\ is reduced to
finding the solution of the generalized eigenvalue problem
\eqn\poleq{
\det\left(H_{ij}-\mu^2N_{ij}\right)=0,\qquad
i,j=0,1,2,\ldots
}
where
\eqn\defkpeq{
  H_{ij}\equiv\left(\basis i,H\basis j\right)\equiv
        K_{ij}+U_{ij},\qquad
  N_{ij}\equiv \left(\basis i,\basis j\right)
}
\def\bet#1{B(#1,#1)}
\eqn\nelem{
  N_{2k,2l}=\bet{2\gamma+k+l+1},\qquad
  N_{2k+1,2l+1}={\bet{2\gamma+k+l+1}\over4\gamma+2k+2l+3}
}
\eqn\kelem{
  K_{2k,2l}=(\beta-1)\bet{2\gamma+k+l},\quad
  K_{2k+1,2l+1}={(\beta-1)\over4\gamma+2k+2l+1}\bet{2\gamma+k+l}
}
\eqn\pelemone{
  U_{2k,2l}=\cases{{(\gamma+k)(\gamma+l)\over2(2\gamma+k+l)}
  \bet{\gamma+k}\bet{\gamma+l}&ff\cr
  {8kl+(8\gamma+1)(k+l)+8\gamma^2+2\gamma\over4( 2\gamma+k+l)}
  \bet{\gamma+k+1/2}\bet{\gamma+l+1/2}
  &bb\cr}
}
\eqn\pelemtwo{
  U_{2k+1,2l+1}=\cases{{(\gamma+k)(\gamma+l)
  \over2(2\gamma+k+l)(2\gamma+k+l+1)}
  \bet{\gamma+k}\bet{\gamma+l}&ff\cr
  {4kl+(4\gamma+1)(k+l)+4\gamma^2+2\gamma\over
  2(2\gamma+k+l)(2\gamma+k+l+1)}
  \bet{\gamma+k+1/2}\bet{\gamma+l+1/2}
  &bb\cr}
}
$H_{ij},N_{ij}=0$ when $i+j\equiv1\ {\rm mod\/}\,2$.
The boundary conditions mentioned in the previous sections
require that $\gamma>0$ which guarantees that the meson masses
and the matrix elements given above  are  finite.

It is illuminating to study the simplest one dimensional case
analytically.
\def\expecp#1{\left\langle{#1}\right\rangle_{\gamma}}
In this case, the meson mass squared is
\eqn\polest{\expecp{H}
  \equiv{H_{00}\over N_{00}}
  =(\beta-1){4\gamma+1\over\gamma}+
  \cases{{\gamma\over4}{B^2(\gamma,\gamma)\over\bet{2\gamma+1}}&ff\cr
  \left(\gamma+{1\over4}\right){B^2(\gamma+1/2,\gamma+1/2)\over
  \bet{2\gamma+1}}&bb\cr}
}
We may vary $\gamma(>0)$ to obtain an upper bound on the meson
mass.
In the ff case, when the quark mass is positive ($\beta>0$) the
problem is well behaved and it was already established that the
meson mass is positive using \useful.
When $\beta=0$ the meson mass is zero.
When $\beta<0$,  $\expecp{H}$ is not only negative but is
unbounded from below.
The situation parallels that of the $1/r^2$ potential in quantum
mechanics; when the coupling is smaller than the critical value,
the problem becomes ill behaved.
In the bb case, this critical coupling is at $\beta=1$;
when $\beta<1$, the meson mass is unbounded from below and is
not acceptable physically.
As a result, the bosonic quark mass needs to satisfy
the condition $\beta\geq1$, which we shall adopt from now on.
When $\beta\geq1$ the meson mass is positive and the problem is
well behaved.
Using this variational approach, we may establish that
$0<\mu^2<\pi^2/4$ for $\beta=1$.
Consequently, it is not possible to obtain a massless bb meson
in this model.
Combined with the QCD mass inequality to be explained below,
this excludes the possibility of obtaining  massless mesons with
boson constituents.

We may use this formulation for the numerical computation for
the spectrum.
In the bf case, it is natural to choose  $\gamma $ to satisfy
the boundary behavior \bbeq.
This formulation is most convenient for the ff case with light
quark masses $\beta\lesssim1$.

The basis we have chosen is not orthonormal. An orthonormal
basis is
\eqn\polortho{
  \basis i'=x^{\gamma_0}(1-x)^{\gamma_1}
G_n(2\gamma_0+2\gamma_1+1,2\gamma_0+1 ;x)
}
where $G_n(a,b;x)$ denotes the $n$--th Jacobi polynomial.
While we may compute the matrix elements of the diagonalization
problem analytically using this orthogonal basis, we have not
found a compact general
formula for the matrix elements as in \poleq\ and we shall
not  use this approach here.
\subsec{Multhopp's method}
Another numerical method we shall employ, which is sometimes
called Multhopp's wing dynamics, is valid for all
quark masses in the three ff, bf and bb cases.
This method has previously been applied to the ff case with
mesons formed from
quarks with identical masses  in  \ref\MULTHOPP{
A.J.~Hanson, R.D. Peccei, M.K. Prasad, \npb{\bf B121} (1977)
477\nl
R.C.~Brower, W.L.~Spence, J.H. Weis, \prd{\bf 19D} (1979) 3024\nl
S. Huang, J.W.~Negele, J.~Polonyi, \npb{\bf B307} (1988) 669\nl
R.L.~Jaffe, P.F. Mende, \npb{Bf 369} (1992) 182}.
Here, we apply this method to a meson made from a quark and
an anti-quark of different mass in the ff, bf and bb cases.
We approximate the meson wave function by
\eqn\multhopp{\tvp(x)=\sumk na_n\sin n\theta,\qquad
  x\equiv{1+\cos\theta\over2}}
Using integration by parts, the integral equation \bseq\ for the meson
wave function reduces to
\eqn\mone{
  \mu^2\tvp=\sumk nM_n(\theta)a_n}
\eqn\mndef{
  M_n(\theta)\equiv2\left({\be a-1\over1+\cos\theta}+
  {\be b-1\over1-\cos\theta}\right)\sin n\theta+2\pi
  \left({n\sin n\theta\over\sin\theta}+B_n(\theta)\right)}
where
\eqn\bdefeq{
  B_n(\theta)\equiv\cases{0&ff\cr
  {\cos\theta\cos n\theta\over 2(1+\cos\theta)}&bf\cr
  -{\cos\theta\cos n\theta}+\delta_{n,1}/8\over
\sin^2\theta&bb\cr}}
We solve the equation by evaluating it at the points
$\theta=\theta_{n},\ n=1,2,\ldots,K$ where we used the notation
$\theta_j\equiv\pi j/(K+1)$.
Using the relation
\eqn\sineq{\sumk
k\sin\theta_{nk}\sin\theta_{mk}={K+1\over2}\delta _{nm}}
the problem reduces to a finite dimensional matrix eigenvalue
problem
\eqn\eigvalprob{\sum_k A_{nk}a_k=\mu^2a_n,\quad
A_{mn}\equiv{2\over K+1}\sumk k\sin\theta_{mk}M_n(\theta_k)}

We now comment on the relative merits of the
numerical schemes we used.
First, we  note that when solving the problem
numerically, it is beneficial both in terms efficiency
and in terms of accuracy to be able to compute
the matrix elements of the diagonalization problem analytically.
If one does not insist on this property, we
may use \useful\ to compute the matrix elements by numerically
performing the double integral for any basis satisfying the boundary
conditions.
Multhopp's method is robust in that it tends to produce matrix
elements of order one even when $K$ is large and does not give
rise to almost singular matrices numerically.
Also, the diagonalization problem is well behaved for all values
of the quark masses.
Except in the region of light
quark mass ($\beta\lesssim1$) the convergence is sufficiently
fast;
$K=200$ is more than enough to obtain the meson mass to eight
digits.
$K=800$ is a reasonable computational task  requiring
the order of  an hour of cpu time on a current workstation.
For the ff case when the quarks are light and are of equal mass,
the basis involving powers of the momentum fraction  is useful.
A ten dimensional diagonalization suffices to obtain the
spectrum to  five significant digits.
However this approach tends to lead to almost singular matrices
numerically as we increase the dimension of the variational space.
It may be possible to overcome this problem by a clever choice
of basis.
The convergence is slower than the Multhopp's method
except in the ff case with light quark masses of equal mass.
\subsec{The meson spectrum and the QCD mass inequality}
In all the three ff, bf and bb cases, the spectrum leads to a linear
Regge trajectory for the higher mass states as we saw in \regge.
Here, we plot the examples for three sets of parameters,  (i)
$\be a=\be b=1$, (ii) $\be a=1,\be b=10$ and (iii)  $\be a=\be
b=10$ for ff, bf and bb cases in \fig\figregge{}.
Here and below, unless otherwise stated, the numerical errors
are too small to be visible on the plots.
\nl
\epsfysize=6.0cm\centerline{\epsfbox{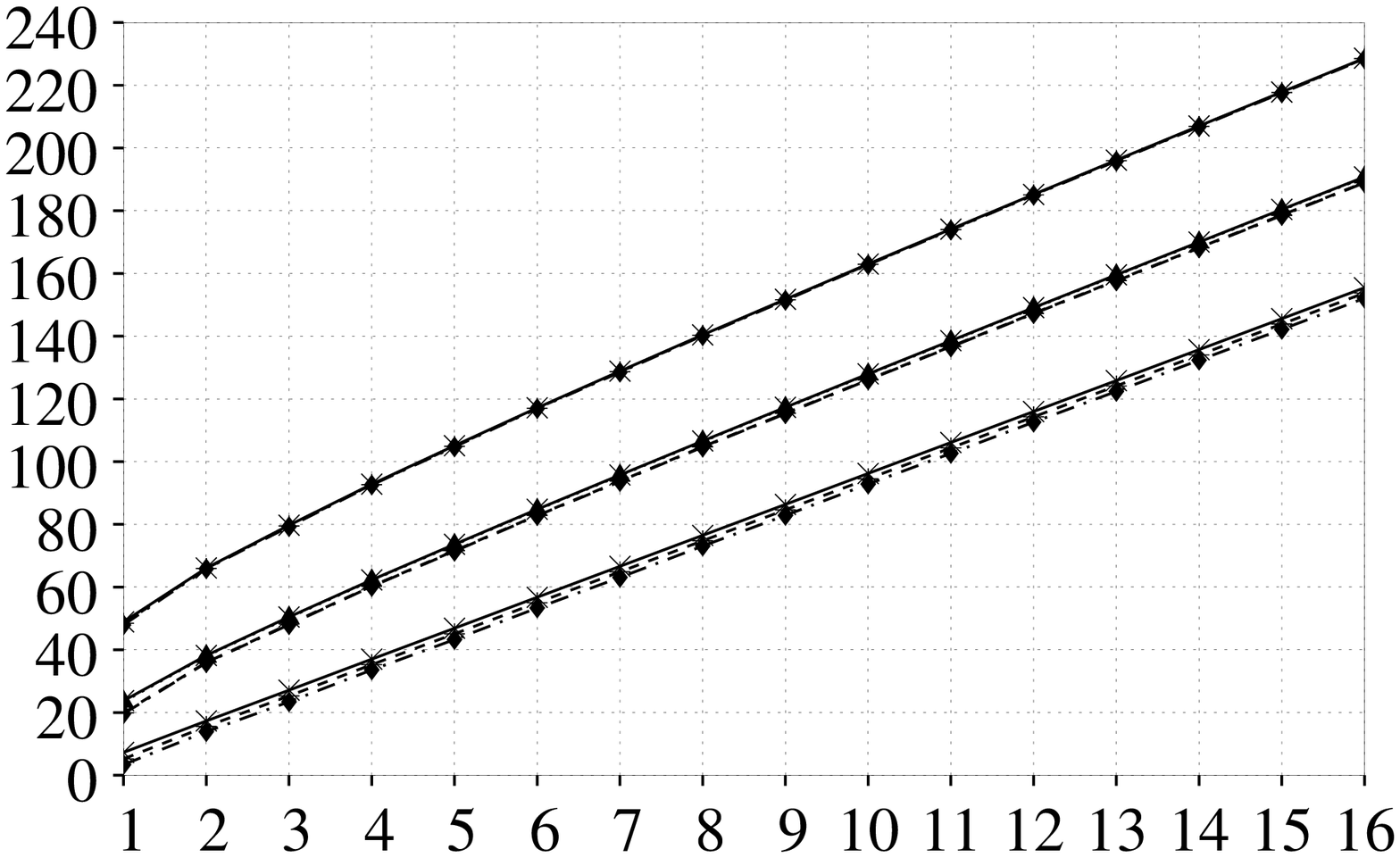}}\nl
\figregge{\ \it The spectrum of mesons for the cases
(i) $\be a=\be b=1$, (ii) $\be a=1,\be b=10$ and (iii)  $\be
a=\be b=10$.
The horizontal axis is labeled by the number when we count from
the lightest meson (``principal quantum number'')  and the
vertical axis is labeled by the meson mass squared counted in
the units of the QCD scale ($g^2\nc/\pi$).
The spectrum for the ff, bf and bb cases are joined by solid,
dashed and dot dashed lines, respectively.
The ff, bf and bb cases are barely distinguishable for (i) and
indistinguishable in the other cases.
}
\medskip
{}Let us now analyze how the QCD inequality is realized in QCD
in $1+1$ dimensions in the large $\nc$ limit.
The QCD mass inequality was originally shown for the four
dimensional case \INEQ\NUSSINOV.
At a formal level, the inequality may be extended to the ff, bb
and bf cases in two dimensions.
Given the analytical form of the masses when the quark masses
are heavy, we may analyze how the QCD mass inequality is
satisfied in this case.
Comparing the masses, we obtain
\eqn\massdiff{2\mu_{ab}-(\mu_{aa}+\mu_{bb})
  ={3\pi^{1/3}\over2(\be a\be b)^{1/6}}
  \left[
  \left(\sqrt{\be a}+\sqrt{\be b}\right)^{1/3}-
   2^{2/3}\left(\be a^{1/6}+\be b^{1/6}\right)
  \right]
   +\c O(\beta^{-1/2})\geq0}
to this order in all three ff, bf and bb cases.
We see that the mass difference arises not at leading order but
at next order since the leading order only contains the rest
masses of the quarks.
Furthermore, we see that the mass inequality can be an equality
to this order only if the masses of the quarks are the
same.
In the  ff and bb cases, this inequality is an exact equality
when the masses of the bound quark and anti-quark are the
same ($\be a=\be b$) since the mesons being compared
can be the same meson.
When applying the mass inequality to the bf case, this is not
the case so that it can be and is a proper inequality.
In the bf case $\mu_{ab}$ in the inequality is the mass of bf
fermionic meson and the $\mu_{aa}$, $\mu_{bb}$ are the masses of
the ff, bb mesons.
When $\be a=\be b(\equiv\beta)$, we may compute to next order
and obtain
\eqn\bfineq{2\mu_{ab}-(\mu_{aa}+\mu_{bb})={1\over\sqrt\beta}
\left(\sqrt2-1-{\pi\over8}\right)\sim{0.022\over\sqrt\beta}>0}
As mentioned above, since the expansion in $\beta^{-2/3}$ is an
asymptotic one, this analytic derivation of the inequality is
not rigorous and perhaps should be considered illustrative.
The numerical results below clearly show that the inequality in
the bf case is a proper inequality.

The QCD mass inequality may be analyzed numerically for
arbitrary values of the quark mass.
We plot  the normalized mass inequality
$\left(\mu_{ab}-(\mu_{aa}+\mu_{bb})/2\right)/\mu_{ab}$, which is
dimensionless,  against $\be b$
in \fig\ineqone{}, \fig\ineqtwo{}
for $\be a=1,10$ respectively.
We immediately see that the QCD mass inequality is a proper
inequality in the bf case.
The values for the normalized mass inequality at $\beta=0$, while
not visible on the plots, are finite and can go up to $0.3$ and
$0.1$ for the $\beta=1$ and $\beta=10$ cases, respectively.
Also,
even though the relative mass inequality is increasing as we
increase $\be b$, at some point, the relative inequality starts to
decrease and  is never more than 0.012, 0.002 for $\be b>\be a$
in the $\beta=1,10$ cases respectively.
For  $\beta=1$ the numerical data can have appreciable errors as
shown on the plots.
The errors were crudely estimated as follows:
Since the numerical data has not completely converged, we
linearly extrapolated the finite dimensional results by using
the analytically known $\be a=\be b=0$ ff case as a guide.
The error in the extrapolation was estimated by the statistical
error in the extrapolation when various sets of data were taken.
\nl
\epsfysize=6.0cm\noindent{\epsfbox{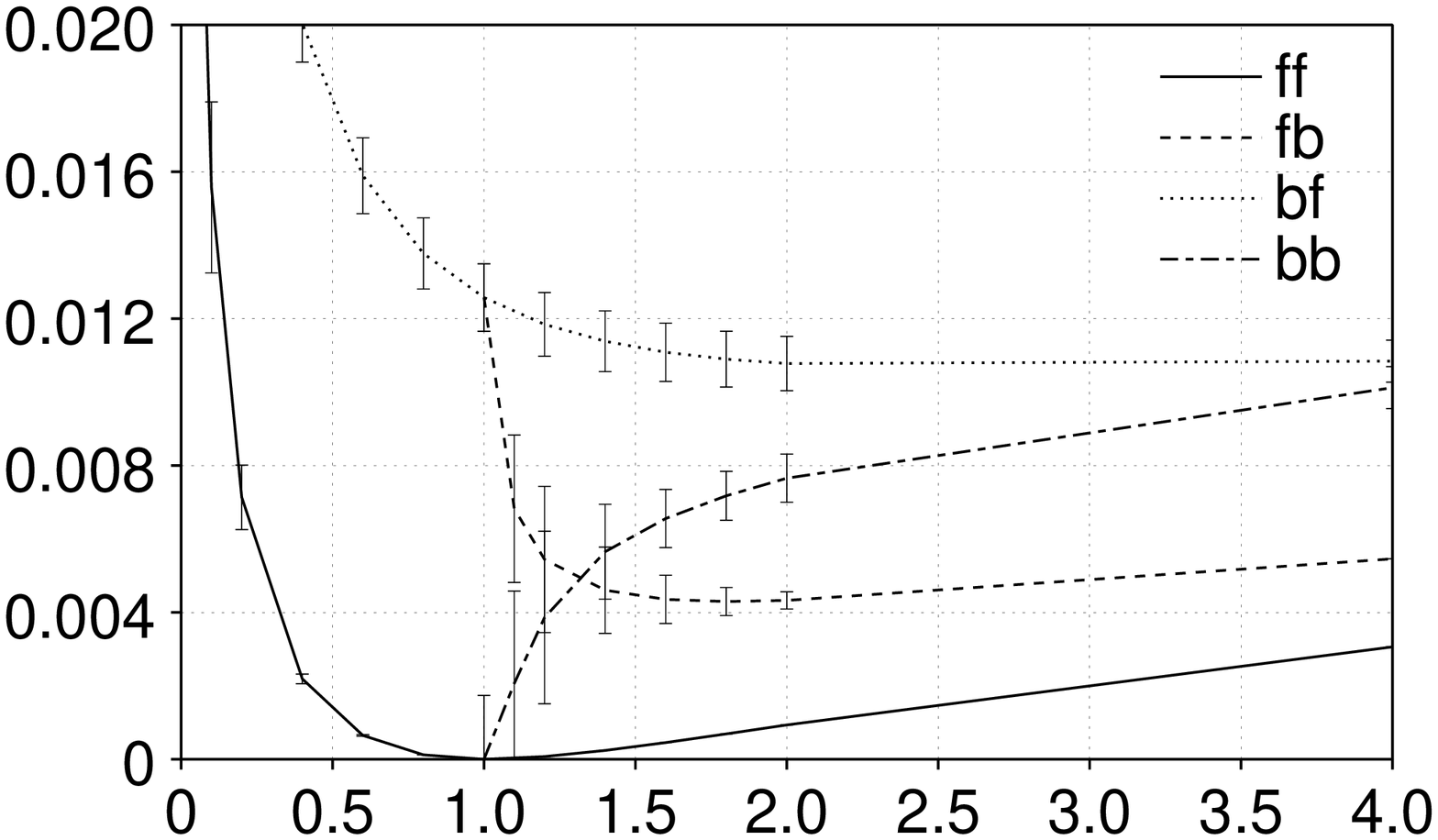}}\nl
\epsfysize=6.0cm{\epsfbox{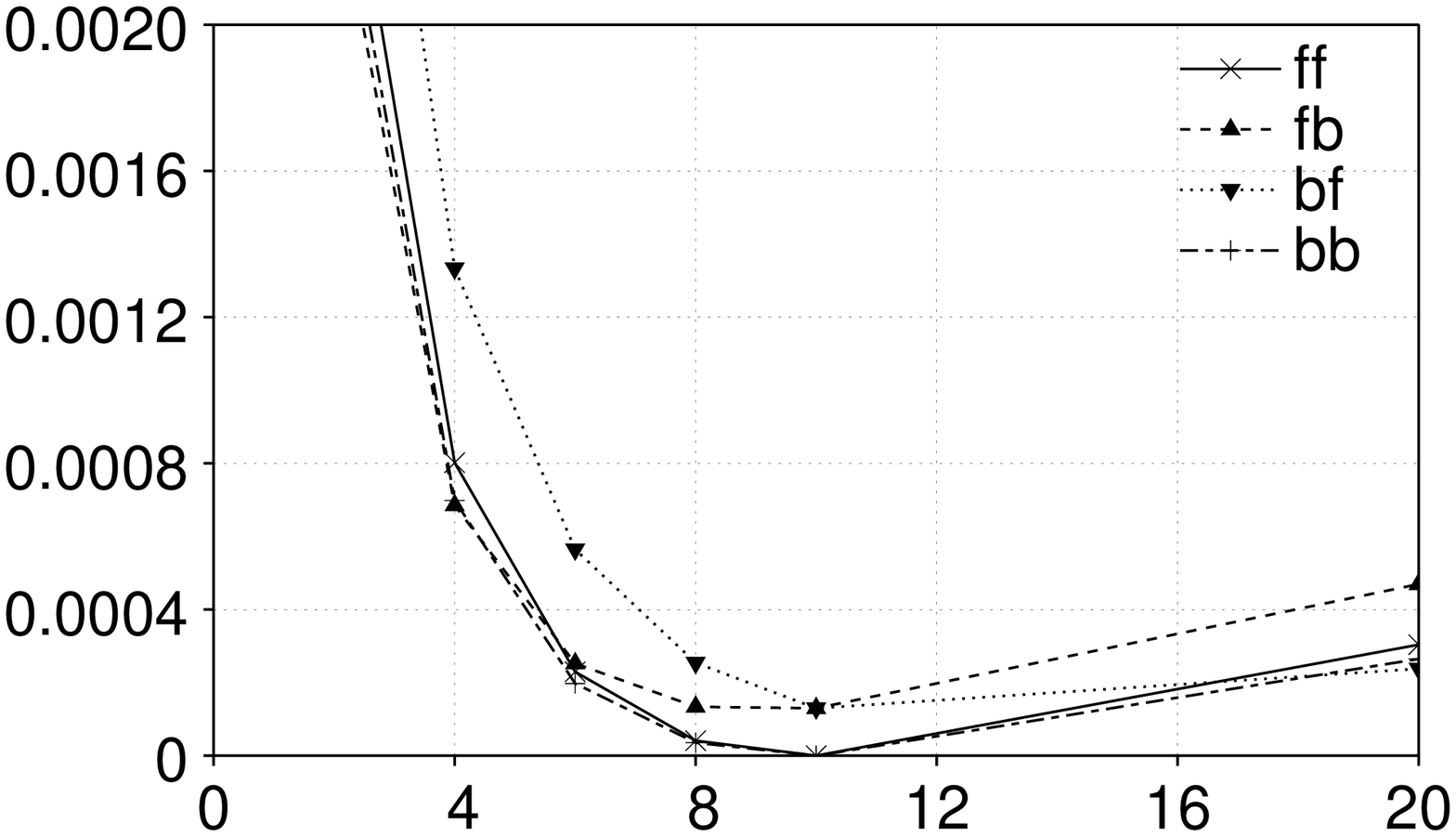}}\nl
\nl\ineqone, \ineqtwo {\ \it
The relative QCD mass inequality for the cases $\be
a=$1 and 10.
fb (bf) means that $\be a$ is the mass of the fermionic
(bosonic) quark.}
\nl

One property that strikes the eye is that the QCD mass
inequality is surprisingly small, even when the quark masses are
of order one in QCD scale.
Here, we remind the reader that the QCD scale is taken to be
$g^2\nc/\pi$, which is a conservative choice when we consider
that the Regge slope parameter is larger by a factor of $\pi^2$
as in \regge.
This near saturation of the inequality is an indication of how
well the constituent quark picture works in this model.
This is in agreement with the general argument  concerning
the validity of the constituent quark
picture in the large $\nc$ limit \ref\WEINB{S. Weinberg,
\prl {\bf 65} (1990) 1177, 1181 and references therein.
}.
It is naturally of interest, then, to compare the constituent
quark mass against the naive or the bare quark mass, which is
done in \fig\constquark{}.
In this figure, $\mu_{aa}/2$  is plotted against the bare quark
mass.
\epsfysize=6.0cm\centerline{\epsfbox{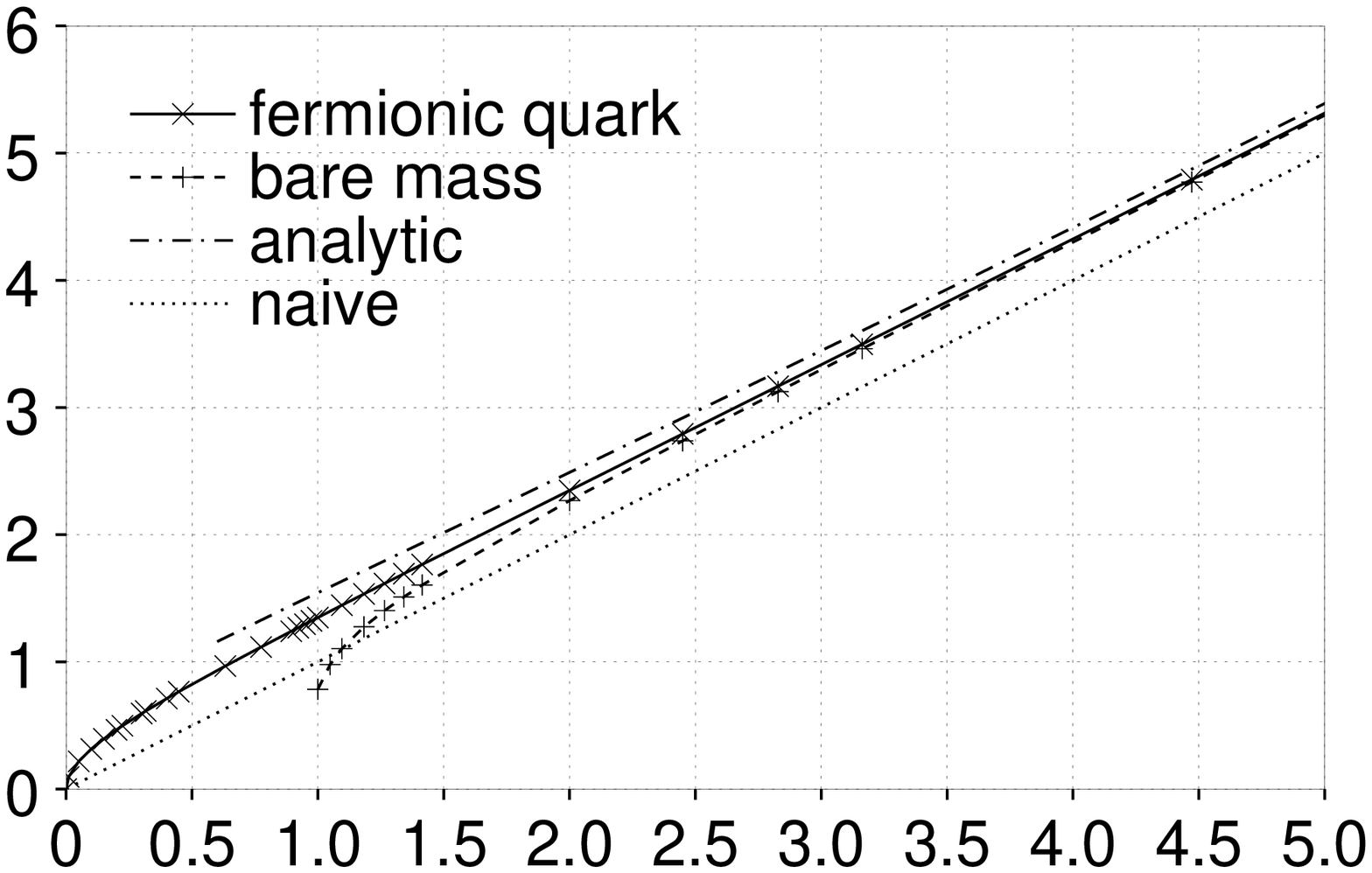}}
\nl\constquark {\ \it The constituent quark masses of the
fermion and the boson quarks  computed from the meson masses
compared against the naive quark mass, $\sqrt{\beta_{f,b}}$.
``analytic'' refers to the formula derived analytically from
\mesonmass.}\medskip\noindent
The constituent quark mass is close neither to the bare quark
mass $m_q$ nor the quantum corrected mass
$\sqrt{m_q^2-g^2\nc/\pi}$,  but is larger than $m_q$ by an
amount of $\c O(1)$ when the mass is of the order of the QCD
scale.
This shows that even though the constituent quark picture seems
valid, the constituent quark is not the naive quark but a
``dressed'' quark.
When the quark mass is heavy, the difference between the
constituent quark mass and the naive quark mass may be
understood from analyzing the non--relativistic linear
potential.\foot{We would like to thank
H. Kawai for pointing this out.}
\newsec{$q\overline q$ hadron vertex, confinement  and
asymptotic freedom}
In this section, we analyze some other physical properties of
the ``bf meson''.
The results are somewhat similar to the ff case \CCG\ and the bb
case \SHEI.
In the bf case, we should note that unlike the ff and the bf
cases, the ``meson'' is a fermion.

First, let us analyze the $q\overline q$ scattering matrix. The
equation satisfied by the $q\overline q$ scattering matrix is
essentially that of the equation satisfied by the meson wave
function \bseq.
First we make use of  the Lorentz structure of the scattering
matrix  and reduce it to its essential scalar component by defining
$  T_{\alpha,\beta}(p,p';r) = (\gamma_-)_{\alpha\beta} T(p,p';r)$.
The reduced matrix element  satisfies the equation
\eqn\teq{
  T(p,p';r)=ig^2L(p_-,p'_-)+2ig^2\nc\intk
  D(k)\hat S(k-r)  L(k_-,p_-) T(k,p';r)}
Here we defined $L(x,y)\equiv(x+y)/(x-y)^2$ and we denoted by
$\hat S(p)$ the component of the fermion propagator that contributes
to the scattering matrix $\gamma_-S(p)\gamma_-\equiv
2\gamma_-\hat S(p)$.
This equation may be solved in a manner closely related to that
of the meson equation \bseq.
Define $\phi(p_-,p;r) $ as
\eqn\phidefeq{
  \phi(p_-,p;r)\equiv\int dp_+\,D(p)\hat S(p-r)T(p,p';r)}
so that
\eqn\phidefeqtwo{
  T(p,p';r)=ig^2L(p_-,p'_-)+{ig^2\nc}
  \int{dk_-\over2\pi}\,  L(k_-,p_-)\phi(k_-,p;r)}
$p_+$ in \phidefeq\ may be integrated out explicitly to obtain
the following equation for $\phi$
\eqn\scatteq{\tilde H\phi(x,y;r)=\mu^2\phi(x,y;r)+
  {\pi^2\over\nc r_-x}L(x,y)}
where $\tilde H$ was defined in \bseq.
We may obtain the solution to this equation using the meson wave
functions as
\eqn\scattsol{\phi(x,x';r)={\pi g^2\over r_-}\sum_k{1\over
   r^2-r^2_k}\int_0^1\!\!dy\,
 \tvp_k(x)\overline{\tvp_k(y')}L(x',y')}
Provided that $\{ \tvp_k\}$ satisfies the following properties:
\item{1.} The ``Schr\"odinger equation'' for the meson:
\eqn\mesoneq{
 \tilde H\tvp_k(x)=\mu_k^2\tvp_k(x)}
\item{2.} Completeness:
\eqn\completeness{x\sum_k\tvp_k(x)\overline{\tvp_k(y)}
  =\delta(x-y)}
\item{3.} Orthogonality:
\eqn\orthogonality{
        \int_ 0^1\!\!\!dx\,x\overline{\tvp_k(x)}\tvp_l(x)
      =\delta_{kl}}
\medskip\noindent
Corresponding properties in the ff case have been proven
rigorously in \ref\PROOF{P. Federbush, A. Tromba, \prd{\bf 15D}
(1977) 2913}.

Using the expression for the reduced scattering matrix
\scattsol\  we may reconstruct the scattering matrix as
\eqn\scattmat{
  T(x,x';r)={ig^2\over r_-}L(x,x')+
  \sum_k{2ir_-\over r^2-r^2_k}\Phi_k(x)\overline{\Phi_k(x')}}
where the $q$--$\overline q$--meson vertex $\Phi_k(x)$ is
defined as
\eqn\vertex{
  \Phi_k(x)={g^2\over2r_-}\left(\nc\over\pi\right)^{1/2}
  \int_0^1\!\!\!dy\,L(x,y)\tvp_k(y)}
We see that the $q\overline q$ scattering matrix may be broken
up into the
$q$--$\overline q$--meson vertex and the single particle poles
corresponding to the mesons which are solutions to the bound--state
equation.
All the intermediate physical states to this order in $1/\nc$ are
mesons and we see that the colored particles  are indeed confined.
At higher orders in $1/\nc$, more complicated intermediate states such
as two particle cuts will appear.

\def\spm#1{F_{#1}^{ab}}
To investigate the properties of the meson, we compute some
matrix elements.
Define  a fermionic operator
\eqn\spmeq{\spm\mu\equiv\sum_i{ \phi^i_a}^\dagger\gamma_\mu\psi^i_b}
where $i$ is the color index.
Denoting the $k$--th meson state as $h_k$, we may obtain some matrix
elements such as
\eqn\matelem{\eqalign{
  \expec {0|\spm-|h_k}&=
  -2i\left(\nc\over\pi\right)^{1/2}\int_0^1\!\!\!dx\,\tvp_k(x) \cr
  \expec {0|\spm+|h_k}&=
  2\left(\nc\over\pi\right)^{1/2}{m_f\over r_-}
  \int_0^1\!\!\!dx\,{\tvp_k(x)\over x} \cr}}

Some correlation functions  of these fermionic ``currents" may also be
obtained.  Define the correlation function
\eqn\mdefeq{
  M_{\mu\nu}(q)\equiv \int
  d^2x\,e^{iqx}\left\langle0\left|T{\spm\mu}^\dagger(x)
  \spm\nu(0)\right|0\right\rangle}
\eqn\corrfn{M_{--}(q)={q_-\over q^2}\sum_k\left(\int_0^1\!\!\!dx\,
 \tvp_k(x)\right)^2
  }
The behavior of this element in the deep inelastic region is
\eqn\dies{M_{--}(q)\sim {q_-\over q^2}\ln{q^2\over m^2}
\qquad{ \rm for}\  \ \ q^2\gg m^2,g^2\nc}
This agrees with the
correlation function obtained in the free theory by computing
the Feynman graph in  \fig\figaf{}.
This is, of course, none other than the statement of asymptotic
freedom in this case.
For comparison, the current matrix elements for the ff and the
bb case are
\eqn\ffbb{M_{--}(q)\sim\cases{
        {q_-\over q^2}&ff\cr
        {q_-^2\over q^2}\ln{q^2\over m^2}&bb\cr
        }
\qquad{ \rm for}\  \ \ q^2\gg m^2,g^2\nc}
\epsfysize=3.0cm\centerline{\epsfbox{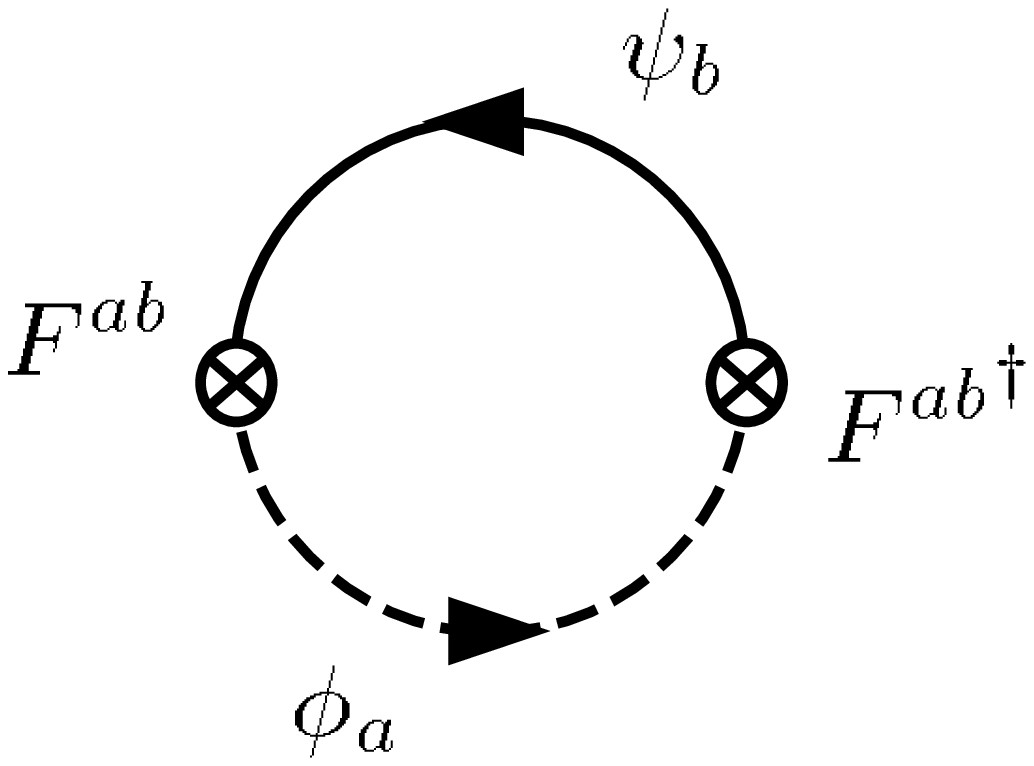}}\nl
\figaf{\ \it Correlation function of the fermionic currents in
the {\rm free} theory.}\medskip\noindent
\newsec{Discussion}
In particle physics and in other areas of physics, such as
condensed matter, we often need to consider particles bound
together by gauge interactions involving boson constituents.  It
is important to have a concrete example where such phenomenon is
analyzed from first principles.
In particular, the precise correspondence between QCD without
quarks and string theory has been recently established in two
dimensions \ref\GT{D.J.~Gross, \npb{\bf B400} (1993) 161\nl
J.~Minahan, \prd{\bf D47} (1993) 3430\nl
D.J.~Gross, W. Taylor, \npb{\bf B400} (1993) 181, {\bf B403}
(1993) 395}.
It is of import to elucidate this correspondence when dynamical
matter is coupled to QCD, putting the previous work on this subject
\THOOFT\ref\STRINGQCD{W.~Bardeen, I.~Bars, A.~Hanson, R.~Peccei,
\prd{\bf D13} (1976) 2364}
on a more rigorous footing.
In this regard, it is crucial to understand in detail the
properties of free strings, which are none other than the mesons
in QCD.
In this paper, we have analyzed the properties of bound states
involving boson constituents both analytically and numerically.
We believe that this concrete model should contribute to the
physical understanding of bound states involving boson
constituents.

There are a few intriguing aspects of our results
which were not anticipated prior to computation:
In mesons involving bosons, we found that it was not possible to
construct massless bound states without additional interaction
other than the gauge interaction.
This is consistent with the result of \ref\BB{W.A.~Bardeen, M.
Bander, \prd{\bf D14} (1976) 2117} where it is found that the
phase transition between the broken symmetry phase and the
symmetric phase in this model is of first order.
This behavior differs from the ff case
where zero mass quarks produced a massless meson.
It is possible that interactions may change some of
these properties.
Interactions that may be naturally added are the $|\phi|^4$
interaction and the Yukawa interaction.
There are, in addition, interaction terms that are
renormalizable in  1+1 dimensions but not in 3+1 dimensions,
such as $(\overline\psi\psi)^2$ and any gauge invariant scalar
self interaction terms.

In this work, we analyzed the QCD mass inequalities in 1+1
dimensions for mesons made from bosons and/or  fermions.
It is important to analyze this non--perturbative inequality in
a model when possible.
We have shown how the inequality works quantitatively in 1+1 dimensions.
An interesting aspect is that the QCD mass inequality is
close to being saturated; as we have seen, to at most few
percent relatively.
Naively, there is no reason to expect that it should be so small
and this is a sign  that the constituent model works well even
when the quark masses are light.
We need to ask if these properties are artifacts of the model we
used, namely the large $\nc$ limit and the fact that the model
is formulated in 1+1 dimensions.
In the large $\nc$ limit, quark loops are suppressed for group
theoretical reasons and one may wonder if this is the cause for
the saturation.
First, even in two dimensional QCD in the large $\nc$ limit,
the inequality is not exactly saturated and we know of no solid
argument short of a concrete calculation that shows that the
inequality is close to being saturated.
Furthermore, in recent studies in 1+1 dimensions,  it has been
shown that even when the quark loops are not suppressed for
group theoretical reasons, the lighter hadrons are very
well approximated by a small definite number of
partons, so that the quark loops are
suppressed dynamically \ref\ADJ{S. Dalley, I.R. Klebanov,
\prd{\bf D47} (1993) 2517\nl
G. Bhanot, K. Demeterfi, I.R. Klebanov,
\prd {\bf D48} (1993) 4980, \npb~{\bf B418} (1994) 15\nl
D. Kutasov \npb{\bf B414} (1994) 33\nl
J. Boorstein, D. Kutasov \npb {\bf B421} (1994) 263}.
While it is unknown whether this feature is preserved in higher
dimensions,
we should keep in mind the successes of the constituent quark
model
\WEINB\ref\LFF{
For instance, see Wilson et al., \prd{D49} (1994) 6720, and
references therein.}.

1+1 dimensional QCD has served particle physics well as a
testing ground for various  properties of QCD.
In this paper, we have  made an effort to present the method for
computing the wave function and the spectrum of ff, bb and
bf bound states in 1+1 dimensional QCD in a concrete manner.
We believe that this will be useful for further investigations in
this field.
\medskip\noindent
{\bf Acknowledgments: }\nl
We would like to thank Eric  D'Hoker, Norisuke Sakai and
Hidenori Sonoda for numerous discussions and encouragement.
We would also like to thank K.~Harada and M.~Taniguchi
for illuminating  discussions and for pointing out the references in
\SMODEL.
This work was supported in part by the Grant in Aid for
Scientific Research from the Ministry of Education, Science and
Culture and the Research Fellowship of the Japan Society for the
Promotion of Science for Young Scientists.
\listrefs
\vfil\eject
\end